\documentclass[twocolumn,prb,aps,endfloats,showpacs]{revtex4}
\usepackage{epsfig}

\begin{document}

\title{Brueckner-Hartree-Fock study of circular quantum dots} 

\author{A. Emperador}
\affiliation{Departament de F\'{\i}sica i Enginyeria Nuclear, 
Campus Nord B4-B5, Universitat Polit\`ecnica de Catalunya,
E-08034 Barcelona, Spain}

\author{E. Lipparini}
\affiliation{Departament ECM, Facultat de F\'{\i}sica,
Universitat de Barcelona, Diagonal 647,
E-08028 Barcelona, Spain}
\altaffiliation{Permanent address: Dipartimento di Fisica, 
Universit\`a di Trento, and INFN sezione di Trento, I-38050 Povo, Italy}

\author{Ll. Serra}
\affiliation{
Departament de F\'{\i}sica, Universitat de les Illes Balears, and
Institut Mediterrani d'Estudis Avan\c{c}ats IMEDEA (CSIC-UIB),
E-07122 Palma de Mallorca, Spain}

\date{October 28, 2005}

\begin{abstract}
We calculate 
ground state energies in the Brueckner-Hartree-Fock theory
for $N$ electrons (with $N\le 20$) confined to a circular quantum
dot and in presence of a static magnetic field. Comparison
with the predictions of Hartree-Fock, local-spin-density
and exact configuration-interaction theories is made. 
We find that the correlations taken into account in 
Brueckner-Hartree-Fock calculations give an important contribution to 
the ground state energies, specially in strongly confined dots.
In this high-density range, corresponding in practice to 
self-assembled quantum dots, the results of Brueckner-Hartree-Fock 
calculations are close to the exact values and better than those 
obtained in the local-spin-density approximation.
\end{abstract}


\maketitle

\section{Introduction}

The Hartree and Hartree-Fock (HF) mean field
approaches have been extensively used in the past to study 
atomic-like properties of semiconductor quantum 
dots such as, e.g., those measured in conductance and capacitance 
experiments.\cite{Mei90,Ash92,Tar96,Aus99}
Symmetry restricted approximations, as well as spin and/or space 
unrestricted HF solutions 
have been analyzed.\cite{Kum90,Fuj96,Mul96,Yan99,Reu03,Ser03}
These methods 
give results that are 
satisfactory for a qualitative understanding of many properties of
these systems. However, comparison with exact configuration-interaction
diagonalization\cite{Eto97,Mak90,Pfa93,Haw93,Pal94,Eza97,Emp05} (CI) 
and Quantum Monte Carlo\cite{Har99,Shu99,Bol96,Mak98,Ped00} (QMC) studies 
show discrepancies in the total energies that are substantial on the 
relevant energy scale. 

The discrepancies become even larger 
in the presence of a constant magnetic field which magnifies the
importance of correlations.   
However, at high magnetic field QMC calculations are 
imprecise due to the fixed phase
approximation and the exact diagonalization techniques require  
a number of configurations which increases exponentially 
with the number of particles, 
making the 
calculation unpractical for more than ten particles.
An alternative method that has been shown to be quite accurate 
for cases in which HF yields broken symmetry solutions is that of 
symmetry restoration by projection,\cite{Yan02} although its complexity 
also increases very rapidly with the number of particles.
The ground state properties of many electrons quantum dots can also be
calculated with local-spin-density approximation (LSDA) and current 
density 
functional methods.\cite{Kos97,Hir99,Fer94} 
These approaches are based 
on energy functionals obtained through fits and interpolations of QMC 
results for 
the spin polarized and unpolarized two-dimensional electron gases that do 
not include a magnetic field
and, moreover, provide a single particle spectrum whose physical
interpretation is not clear. Nevertheless,  
LSDA has given good results for quantum dots with low electronic 
densities, 
having Wigner-Seitz radius $r_s \gtrsim 1$. These are the typical 
densities of 
quantum dots made by etching\cite{Tar96} or by lithography,\cite{Mei90} 
usually over a GaAs substrate. 
Another class of quantum dots are the self-assembled 
ones,\cite{Mil97,War98} 
with high electronic densities ($r_s < 0.5$) and great practical interest 
due 
to technological applications such as, e.g., in semiconductor 
lasers.\cite{lasers}
They are usually formed over InAs substrates, have 
small diameters and contain strongly confined electronic states. 
For systems so distant from the homogenous conditions one should not 
expect the LSDA to give accurate predictions and, indeed, most 
calculations on InAs dots have been made using non-DFT-based 
schemes.\cite{Sza00,Haw03,Bro03} 

In this paper we employ the Brueckner-Hartree-Fock (BHF) 
method\cite{PB82,FW71}
to describe electrons in quantum dots
because it
yields good ground state energies at a computational cost that grows 
in a relatively modest way with the number of particles, 
as compared to CI and QMC (in its most accurate versions such as
diffusion QMC). 
This method has been extensively applied in the past
to study nuclei\cite{Bru58} and metallic clusters.\cite{Lip94}
Also, a related scheme based on Bethe-Goldstone equations
has been used to study small excitonic complexes
in quantum dots.\cite{Per00} The underlying assumption of BHF 
theory is a description in terms of independent 
pairs of particles. For each pair, interaction affects the way in 
which the two particles scatter, while the rest of particles are 
considered
noninteracting and their influence is only through the Pauli exclusion 
principle. It is well known that the exclusion principle 
imposed by the Fermi sea
induces a  
modification of the pair wave function at short interparticle 
distances, inducing the so-called  
``wound'' in the wave function. Therefore, BHF theory
describes short range statistical correlations and will 
miss, by construction, long range correlations associated with 
collective motions.\cite{PB82,FW71}

Here we compare BHF energies for electrons in quantum dots 
with the available results from methods attempting a direct solution of
the many body Schr\"odinger equation like CI diagonalization
and QMC, as well as with approximate models such as LSDA and HF. 
We find that BHF is always more accurate than HF. 
It recovers 
the exact result in the strong-confinement limit and at high
magnetic fields, when the maximum-density-droplet (MDD) is formed,
it yields 
more accurate energies than the LSDA ones.  
Furthermore, in the case of electronic 
densities typical of self-assembled quantum dots, BHF 
also gives much better results than LSDA even in the absence of magnetic 
fields. 

\section{The BHF scheme}

This section recalls some essential points of the BHF theory
deriving several relevant equations and quoting others without proof.
The restriction to circular symmetry solutions and its practical
implications are also discussed.

\subsection{Theory}

Let us consider a system of $N$ electrons with Hamiltonian 
\begin{equation}
\label{eq1}
H=\sum_i^N h_0(i) + \sum_{i<j}^N v_{ij}\; ,
\end{equation}
where
$h_0=t_i+v_{ext}(r_i)$, $t_i$ is the single particle kinetic energy
which can include the effect of the magnetic field,
$v_{ext}$ the confining potential  and
$v_{ij}=e^2/\epsilon|{\bf r}_i-{\bf r}_j|$, with $\epsilon$ the 
dielectric constant.
The 
ground state $\vert\Psi\rangle$
is solution of the many-body Schroedinger equation 
\begin{equation}
\label{eq2}
(H-E)\vert\Psi\rangle=0\;,
\end{equation}
where $E$ is the ground state energy. Let us now consider an
independent particle model (the HF model
in the following) where the
eigenstates $\vert\Phi_n\rangle$ are Slater determinants solutions of the 
equation
\begin{equation}
\label{eq3}
(H_{HF}-W_n)\vert\Phi_n\rangle=0\; ,
\end{equation}
where $H_{HF}={\cal C}+\sum_i^N(h_0(i)+U_i)$, with $U_i$ the HF  
potential and ${\cal C}$ a constant. 
The HF ground state determinant and energy are given
by $\vert\Phi_0\rangle\equiv|HF\rangle$ 
and $W_0\equiv E_{HF}$, respectively.

The residual interaction $V_{\it res}$ fulfills
\begin{eqnarray}
H&=&H_{HF}+V_{\it res}\label{eq4}\; ,\\
V_{res}&=&\sum_{i<j}^N v_{ij}- \sum_i^N U(i) - {\cal C}\; .
\end{eqnarray}
HF theory yields the following general matrix elements 
\begin{eqnarray}
\langle HF|V_{\it res}|HF\rangle &=&0\; ,\label{me1}\\
\langle HF|V_{\it res}|mi^{-1}\rangle &=&0\; ,\label{me2} \\
\langle HF|V_{\it res}|mni^{-1}j^{-1}\rangle &\ne& 0\; ,\label{me3}
\end{eqnarray}
where we have used the standard notation for particle-hole (ph)
excitations, i.e., indexes $i$, $j$ ($m$, $n$) refer to
orbitals below (above) the Fermi energy
and $|mi^{-1}\rangle$ is the Slater determinant obtained
promoting one electron from orbital $i$ to orbital $m$
in $|HF\rangle$. Note that only
2p-2h excitations yield non vanishing transition matrix elements
since the two-body nature of $V_{\it res}$ ensures that matrix 
elements between determinants differing in more than two orbitals 
will again vanish.
Another immediate consequence from Eqs.\ (\ref{me1})-(\ref{me3})
is that $E_{HF}=\langle HF| H | HF\rangle$. 

We can write
\begin{equation}
\label{eq5}
\vert\Psi\rangle=|HF\rangle +\sum_{n\ne0}a_n\vert\Phi_n\rangle~~.
\end{equation}
From Eqs. (\ref{eq2}),(\ref{eq4}),(\ref{eq5})
one easily finds that
\begin{equation}
\label{eq6}
(H_{HF} -E)\left(|HF\rangle +\sum_{n\ne0}a_n\vert\Phi_n\rangle \right)
+V_{\it res}\vert\Psi\rangle=0\; .
\end{equation} 
Multiplying Eq.\ (\ref{eq6}) by $\langle HF|$ on the left one gets
\begin{equation}
\label{eq7}
E=
E_{HF}+ \langle HF| V_{\it res} | \Psi\rangle~~.
\end{equation}
If multiplying by $\langle\Phi_n|$
one finds $a_n={\langle\Phi_n|V_{\it res}|\Psi\rangle\over E-W_n}$ and 
hence 
the following implicit equation is obtained 
\begin{equation}
\label{eq8}
|\Psi\rangle = |HF\rangle+\sum_{n\ne0}
\frac{\langle\Phi_n|V_{\it res}|\Psi\rangle}{E-W_n}|\Phi_n\rangle~.
\end{equation}
This equation
can be solved by iteration taking as starting energy the HF one:
\begin{equation}
\label{eq9}
|\Psi\rangle = |HF\rangle +
\sum_{n\ne 0}\frac{\langle\Phi_n|V_{\it res}|HF\rangle}
{E_{HF}-W_n}|\Phi_n\rangle+\cdots~,
\end{equation}
yielding for the energy
\begin{equation}
\label{eq10}
E= E_{HF}+ 
\sum_{n\ne 0}\frac{|\langle\Phi_n|V_{\it 
res}|HF\rangle|^2}{E_{HF}-W_n}+\cdots~.
\end{equation}
At the first order in $V_{\it res}$ this equation gives a result for
the energy which coincides with the one of first order perturbation
theory, summing all the orders we get the correlation energy
in the ladder approximation. This is more clear
defining the $G$-matrix by the relation
\begin{equation}
\label{eq11}
G|HF\rangle=V_{\it res}|\Psi\rangle~.
\end{equation}
We then get the Bethe-Goldstone implicit equation for $G$:
\begin{equation}
\label{eq12}  
G=V_{\it res} +\
\sum_{n\ne0} V_{\it res}
\frac{|\Phi_n\rangle\langle\Phi_n|}{E-W_n}G~~,
\end{equation}
and from Eq. (\ref{eq7})
\begin{eqnarray}
\label{eq13}
E&=&E_{HF} + \langle HF|G|HF\rangle\nonumber\\
&=& E_{HF} + \sum_{n\ne0}
\frac{\langle HF|V_{\it res}|\Phi_n\rangle\langle\Phi_n|G|HF\rangle}
{E-W_n}~.
\end{eqnarray}
Only 2p-2h determinants yield a non-vanishing contribution 
to the sum of Eq.\ (\ref{eq13}), which can thus be reduced to
a sum of two-body matrix elements. Assuming $E=E_{HF}$ on
the right-hand-side, as in the ladder approximation, 
one has
\begin{eqnarray}
\label{eq14}
E &=& E_{HF} + 
\frac{1}{2}
\sum_{ijmn}\frac{\langle ij|v|mn\rangle}
{\epsilon_i+\epsilon_j-\epsilon_m-\epsilon_n} \times\nonumber\\
&& \qquad\qquad\qquad \left( \langle mn|g|ij\rangle 
- \langle mn|g|ji\rangle\right) ~,
\end{eqnarray}
where the $\epsilon_{\alpha}$ are the HF single particle energies
and we have associated the $G$ matrix with an effective two-body 
interaction $g$.

In order to have a practical computational scheme it remains 
now to specify the two-body matrix elements of $g$ in 
Eq.\ (\ref{eq14}). This is accomplished within the BHF 
independent-pair model,\cite{PB82}
where the off-diagonal matrix elements are
found from
\begin{eqnarray}
\label{eq15}
\langle mn|g|ij\rangle &=& \langle mn|v|ij\rangle\nonumber\\
&+& \sum_{pq}\frac{\langle mn|v|pq\rangle\langle pq|g|ij\rangle}
{\epsilon_i+\epsilon_j-\epsilon_p-\epsilon_q}\; .
\end{eqnarray}
The ground state energy of Eq. (\ref{eq14}) with the matrix elements 
obtained from Eq. (\ref{eq15}) is the BHF energy which sums all
the ladder diagrams corresponding to the iterated solutions of Eq. 
(\ref{eq10}). 

\subsection{Circular symmetry restriction}

In this work we restrict to circular symmetry cases, where
the HF orbitals can be factorized as
\begin{equation}
\label{eq16}
\langle {\bf r}\sigma|i\rangle 
\equiv R_{n_im_i}(r) {e^{im_i\theta}\over\sqrt{2\pi}} 
\chi_{\mu_i}(\sigma)\; , 
\end{equation}
where $n_i=0,1,\dots$, $m_i=0,\pm1,\dots$, and $\mu_i=\pm1/2$ are the 
principal, angular
momentum, and spin quantum numbers, respectively.  
In this situation the angular and spin parts of the two-body matrix 
elements 
yield selection rules on the corresponding quantum numbers and
the matrix elements reduce to 
\begin{eqnarray}
\label{eq17}
\langle ab|v|cd \rangle &=&
\delta_{m_a+m_b,m_c+m_d}\,\delta_{\mu_a\mu_c}\,\delta_{\mu_b\mu_d}\times\nonumber\\
&& I_r(R_{n_am_a},R_{n_bm_b},R_{n_cm_c},R_{n_dm_d})\; ,
\end{eqnarray}
where $I_r$ is a radial integral that we compute numerically. 
Note that through Eq.\ (\ref{eq15}) the same angular momentum selection 
rules apply to $\langle ab|g|cd\rangle$ and that both matrix elements are 
real.

The two-body matrix elements of $g$ required for the
evaluation of the total energy, Eq. (\ref{eq14}), are found
by solving Eq.\ (\ref{eq15}) as a linear system for
the unknowns $\langle mn|g|ij\rangle$. For each pair $ij$ we have
an independent linear system and the above mentioned selection
rules are very important since they allow a big reduction in the number 
of effectively coupled equations. Since the space of
particle states must be truncated, the convergence of the calculation
with the number of empty HF states has to be controlled. Another 
check of the numerical accuracy must be done regarding the number
of radial points used in the evaluation of the integrals $I_r$
of Eq.\ (\ref{eq17}).  

\section{RESULTS}

Table I compares the energies of $B=0$
ground states of $N$-electron dots in BHF with the results of HF, 
LSDA (using the Tanatar-Ceperley parametrization for the correlation 
energy\cite{Tan89}) and diffusion QMC.\cite{Ped00}
The external confinement is taken of parabolic type 
$v_{ext}(r)=m\omega_0^2 r^2/2$, 
with $m$ the electron effective mass. We refer all energies to the 
confinement energy $\hbar\omega_0$ and characterize the interaction 
strength
by the repulsion-to-confinement ratio $R$, defined as
\begin{equation}
R\equiv \frac{e^2/(\epsilon\ell_0)}{\hbar\omega_0}\; ,
\end{equation}
with $\ell_0$ indicating the oscillator length 
($\hbar\omega_0=\hbar^2/m\ell_0^2$).
The results in Table I correspond to $R=1.89$.
Taking, for instance, typical GaAs values 
$\epsilon$=12.4 and $m^*=0.067 m_e$ the chosen $R$ value 
would correspond to a confinement energy of 
$\hbar\omega_0=3.32$~meV, which reproduces
the experimental value of Ref.\ \onlinecite{Tar96}.
The number of electrons is varied from $N=2$ to 13. 

\begin{figure}[!]
\centerline{\psfig{figure=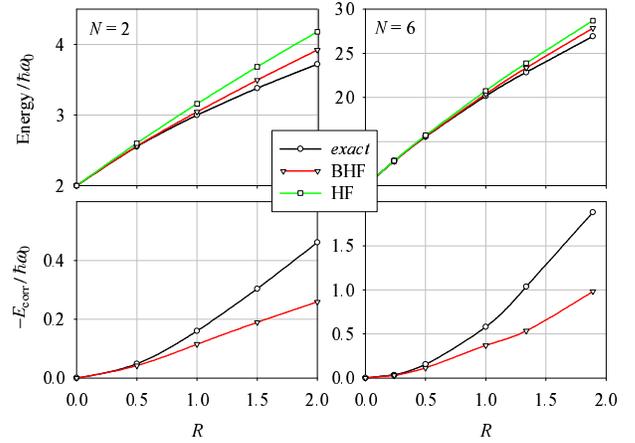,width=3.2in,clip=}}
\caption{
(color online) Upper panels: total energies in different methods
for 2 and 6-electron dots as a function of the 
interaction-to-confinement ratio (see text). 
Lower: Correlation energies within
each model for the same two dots.}
\label{fig1}
\end{figure}

At $B=0$ the BHF energies obviously improve the HF ones, although they are 
still appreciably higher than the QMC and LSDA values. This is
due to the fact that in BHF long-range collective 
correlations are missed. The importance of short range correlations
is expected to increase as the system is more tightly confined, for a 
fixed
strength of the Coulomb repulsion and, thus, a better performance 
of BHF is expected when increasing the confinement strength.
Indeed, this is shown to be 
the case in Fig.\ 1, where for $N=2$ and 6 at $B=0$ we have varied 
the value of $R$.
The lower panels of this Figure display the correlation energies, defined 
as usual by 
subtracting the HF value $E_{HF}$ from the total energy $E$, i.e., 
$E_{\it corr}= E - E_{HF}$.
Note that although the correlation 
energy is globally reduced when
the ratio decreases BHF accounts for a larger part of it.
Actually, for $N=2$ BHF accounts for 71\% of the full (CI) correlation 
energy when 
$R=1$ and 87\% when $R=0.5$. The corresponding figures for $N=6$ are
64\% ($R=1$) and 75\% ($R=0.5$). These numerical results are thus showing
that in the limit of small $R$ BHF converges to the exact correlation 
energy.
It can be also seen from Fig.\ (\ref{fig1}) that at a given value of $R$, 
BHF correlations for the $N=2$ dot are a somewhat larger 
piece of the total correlation energy than for $N=6$. We attribute this
difference to a more important role of the collective effects leading
to long-range correlations not included in BHF for the 6-electron dot. 

\begin{figure}[!]
\centerline{\psfig{figure=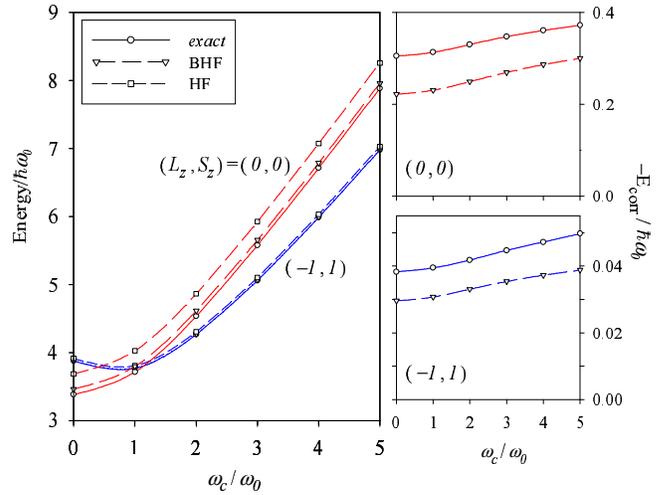,width=3.4in,clip=}}
\caption{(color online) 
Left (right) panels show total (correlation) energies 
for the models and states indicated by the corresponding labels.
The results correspond to the $N=2$ dot in a magnetic field, shown as a 
function 
of the cyclotron frequency (in units of $\omega_0$). 
The interaction-to-confinement ratio $R$ (see text) 
is chosen as $R=1.5$. }
\label{fig2}
\end{figure}

We focus next on the influence of a magnetic field and how
the BHF energies are affected by it. In the symmetric gauge, a 
magnetic field in the $z$ direction (perpendicular to the dot plane) 
induces a modification of the effective confinement from 
$\omega_0$ to $\Omega=\sqrt{\omega_0^2+\omega_c^2/4}$, with 
$\omega_c$ the cyclotron frequency. For increasing magnetic fields
one then expects short range correlations to be enhanced due to the 
stronger confinement and, 
therefore, an improved performance of the BHF method.
Figure \ref{fig2} displays the evolution of the total energy with the 
ratio $\omega_c/\omega_0$
for a fixed value of $R=1.5$ and for a dot with 2 electrons. 
Note that the correlation energy is about an order of magnitude
higher for the singlet than for the triplet.\cite{Pfa93}
As expected, for increasing
values of $\omega_c/\omega_0$ BHF is accounting for a higher
part of the correlation energy, although the increase is rather
moderate. For the singlet state BHF correlations range from 73\% 
to 81\% of the total correlation energy when $\omega_c$ goes
from 0 to $5\omega_0$. The evolution is even flatter for the
triplet, where BHF correlations remain at $\approx 75$\% 
for all values of $\omega_c$. Note also that due to the sizeable
energy correction for the singlet, the singlet-triplet transition 
point is remarkably improved in BHF with respect to HF. 

\begin{figure}[!]
\centerline{\psfig{figure=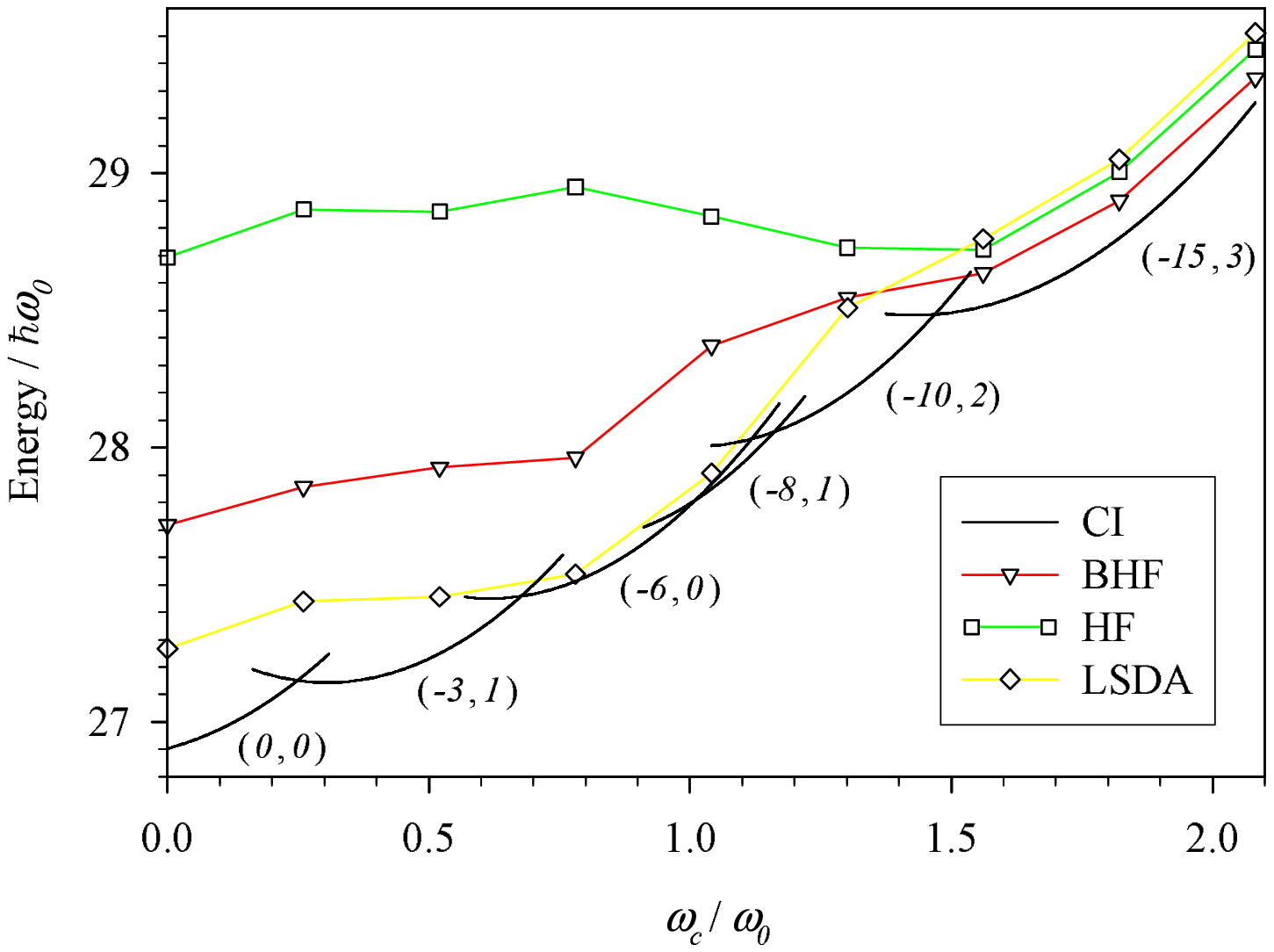,width=3.in,clip=}}
\caption{(color online) Evolution of the total energy of a $6$-electron 
dot
with the magnetic field. The different phases are indicated by
the angular momentum labels $(L_z,S_z)$. A fixed value
of the parameter $R=1.89$ has been used (see text).}
\label{fig3}
\end{figure}

Figure \ref{fig3} shows the evolution with magnetic field 
of the results for a 6-electron dot with a fixed 
interaction-to-confinement ratio of $R=1.89$. 
As compared to the $N=2$ case, this dot shows a much richer phase
diagram, with large variations in ground state angular 
momentum and spin when increasing the magnetic field. 
Most remarkable is the comparison of BHF and LSDA energies: while
LSDA is clearly superior to BHF at low fields the situation
is reversed when entering the fully polarized phase
corresponding to the maximum-density-droplet (MDD). In the MDD 
region the LSDA energy is actually higher than the HF one
and only BHF is able to provide an energy correction in the 
right direction with respect to HF; total BHF energies being 
approximately halfway of CI and HF.

\begin{figure}[!]
\centerline{\psfig{figure=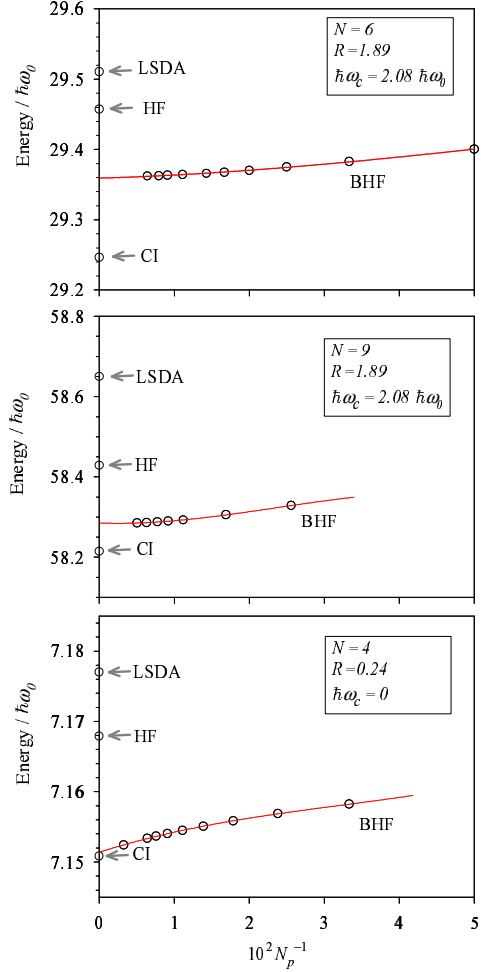,width=2.5in,clip=}}
\caption{(color online) Evolution of the total BHF energy
with the number $N_p$ of empty HF states included in the solution 
of Eq.\ (\ref{eq15}). Each panel shows the 
results for a different quantum dot. The solid line is a cubic fit, 
in powers of $1/N_p$,
allowing extrapolation to the $N_p\to\infty$ limit.
The LSDA, HF and CI energies for each case 
are also shown for comparison.}
\label{fig4}
\end{figure}

All the BHF results shown above have been obtained using 
a large enough space of empty HF states, always checking that 
for the given accuracy convergence in Eq.\ (\ref{eq14}) has been
achieved. In practice, we include the lowest $N_p$ particle
states and repeat the calculation increasing this number.
Figure \ref{fig4} shows the evolution of the BHF energy
with $N_p$, on scale proportional to $1/N_p$,
for three selected cases: the two upper panels correspond to 
six and nine electrons with the MDD configuration in strong magnetic field
and a moderate confinement, while the lower panel shows the case of
four electrons in very strong 
confinement and zero field. The results have been fitted
with a polynomial including powers up to $1/N_p^3$.
In the chosen scale, the $N_p\to\infty$ limit
is given by the intersection of the polynomial fit with the
left vertical axis. It can be seen from Fig.\ \ref{fig4} that the 
convergence
is somewhat faster for the moderate confinement
cases. However, in both examples the evolution with space dimension
is quite smooth, indicating that the correlation energy builds up 
by gathering contributions from many states.
The $N=9$ results of Fig.\ \ref{fig4} provide additional support to our 
preceding conclusion from Fig.\ \ref{fig3} that BHF performs much better 
than LSDA in the MDD region.

As already mentioned, for strong confinements the BHF energies are close
to the exact values. Indeed, the $N=4$ results of Fig.\ \ref{fig4} are 
very
illustrative in this respect since the extrapolated BHF energy 
essentially coincides with the CI result. This strongly confined
system mimics a self-assembled InAs dot with 
$\hbar \omega_0=50$ meV. Following Ref.\ \onlinecite{Bro03} we take for 
this material $m^*=0.024m_e$ and $\epsilon=15.15$  
giving, for $N=4$, a small Wigner-Seitz radius of $r_s\sim 0.12$.

\begin{figure}[!]
\centerline{\psfig{figure=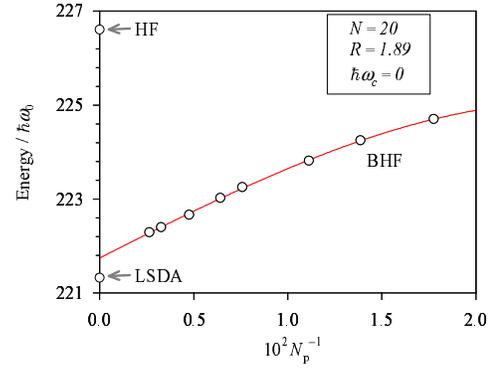,width=2.5in,clip=}}
\caption{(color online) Same as Fig.\ \ref{fig4} for a dot with 
20 electrons and the additional parameters given in the inset.}
\label{fig5}
\end{figure}

To emphasize the possibility of calculating the energies of larger systems 
in BHF
theory we end this section by showing in Fig.\ \ref{fig5} the results 
for an $N=20$ dot. For this number of electrons exact methods
like CI or QMC become extremely demanding and we have not attempted
to compare with them. The evolution of the BHF energy with the number
of particle states resembles that of Fig.\ \ref{fig4} although,
as one could expect, 
larger values of $N_p$ need to be considered for a similar degree of
convergence. The proximity of the extrapolated-BHF and LSDA energies
for $N=20$ is a bit surprising since, as shown in Fig.\ \ref{fig3},
for 6 electrons in the same confinement the difference is larger.
BHF correlations are thus a more important contribution for
$N=20$ than for $N=6$. This can be understood as 
a different degree of {\em magicity} for these two dots. Indeed, a highly 
magic
system is characterized by a distribution of single-particle orbitals
whose energies group in bunches corresponding to quasi-degenerate shells,
with large energy gaps between the shells. 
One expects a quenching of independent-pair motions in a highly 
magic system, with respect to collective motions, and, therefore, 
a relatively worse performance of the BHF theory for them.
Note also that since the $20$-electron dot 
has a higher density than the $6$-electron dot, for the same confinement,
a better performance of BHF theory in the former agrees with our preceding
results regarding the high density limit.

\section{Conclusions}

We have explored the prediction of BHF theory for the correlation 
energy in two-dimensional parabolic dots.
Rather than attempting systematic calculations we intend to 
point out characteristic trends when applying a well established
many body approach such as the BHF theory to parabolic quantum dots.
By comparing with exact calculations we have quantitatively
discussed the relevance of BHF correlations 
as a function of the confinement potential and applied magnetic
field for several quantum dots. 
BHF theory converges to the exact correlation energy 
in the limit of strong confinement potential (small $R$ parameter, high 
densities).
Also relevant is the limit of strong magnetic fields where 
the MDD is formed. In these two regions the standard LSDA is shown to be 
grossly inadequate while BHF stands as a competitive method that sizeably 
improves on HF theory. 
In general, the BHF correlation energy gathers contributions
from many empty HF states, as evidenced by the smooth
convergence with the space dimension. 

Possible extensions of the calculations presented here 
can consider i) relaxing the circular symmetry constraint, and
ii) including self-consistency in the BHF single particle orbitals.
We shall address the latter by 
finding the improved mean-field proposed by Bethe, 
Brandow, and Petschek\cite{BBP} that actually implies the solution
of a double selfconsistency problem, on orbitals and effective 
interaction.
Work along these lines is now in progress.

\acknowledgments 

This work was supported by projects PRIB-2004-9765 and FIS2005-02796 
(Ll.S.).
A. E. acknowledges support from the Generalitat de Catalunya. E.L.
has been supported by DGU (Spain) grant SAB2004-0091.

\begin{table}[!]
\label{tab.ground}
\caption[]{Ground state energies for the dots 
with $2\leq N\leq13$ computed by HF, BHF, LSDA, QMC 
and CI methods.
The energies $E$ in units of
the confinement energy $\hbar\omega_0$ are tabulated as 
$E'=E/\hbar\omega_0$.
A fixed value $R=1.89$ of the interaction-to-confinement ratio
has been used.}
\begin{ruledtabular}
\begin{tabular}{rrrrrr}
$N$ &${E'_{\rm HF}}$ & $E'_{\rm BHF}$ 
    & $E'_{\rm LSDA}$ & $E'_{\rm QMC}$ & $E'_{\rm CI}$\\
\hline
 2 &  4.078 & 3.832& 3.739& 3.650 & 3.646\\
 3 &  8.589 & 8.289& 8.082& 7.979 & 7.957\\
 4 &  13.94 & 13.63& 13.16& 13.26 & 13.06\\
 5 &  20.96 & 20.38& 19.91& 19.76 & 19.53\\
 6 &  28.70 & 27.72& 27.27& 27.14 & 26.82\\
 7 & 37.46  &36.61 &35.96 &35.86 \\
 8 & 46.93  &46.11 &45.46 &45.32 \\
 9 & 57.89  &56.71 &55.79 &55.64 \\
10 & 69.29  &67.75 &67.00 &66.86 \\
11 & 81.54  &79.86 &78.96 &78.86 \\
12 & 94.82  &94.36 &91.71 &91.64 \\
13 & 108.64 &106.36&105.50&105.32\\
\end{tabular} 
\end{ruledtabular}
\end{table}


\printtables
\printfigures

\end{document}